\documentclass[corporate]{adqgem}

\newcommand{\scalt}{\scal{t}}
\newcommand{\scalG}{\scal{q}}
\newcommand{\scalr}{\scal{r}}
\newcommand{\scalh}{\scal{h}}
\newcommand{\scalu}{\scal{u}}
\newcommand{\scalap}{\scal{a_\parallel}}

\newcommand{\vectr}{\vect{r}}
\newcommand{\vecta}{\vect{a}}
\newcommand{\vectu}{\vect{u}}
\newcommand{\vectj}{\vect{j}}
\newcommand{\vectb}{\vect{b}}
\newcommand{\vectw}{\vect{w}}
\newcommand{\vectD}{\vect{D}}
\newcommand{\vecte}{\vect{e}}
\newcommand{\vectz}{\vect{z}}
\newcommand{\vectf}{\vect{f}}
\newcommand{\vectg}{\vect{g}}
\newcommand{\vecth}{\vect{h}}
\newcommand{\vectH}{\vect{H}}
\newcommand{\vectB}{\vect{B}}
\newcommand{\vectE}{\vect{E}}
\newcommand{\vectzero}{\zvect}

\newcommand{\murep}{\mu}
\newcommand{\Pirep}{\Pi}
\newcommand{\chirep}{\chi}
\newcommand{\phirep}{\phi}
\newcommand{\psirep}{\psi}
\newcommand{\etarep}{\eta}
\newcommand{\betrep}{\beta}
\newcommand{\Gamrep}{\Gamma}
\newcommand{\alprep}{\alpha}
\newcommand{\kaprep}{\kappa}
\newcommand{\omerep}{\omega}
\newcommand{\Omerep}{\Omega}
\newcommand{\sigrep}{\sigma}
\newcommand{\delrep}{\delta}
\newcommand{\Delrep}{\Delta}
\newcommand{\gamrep}{\gamma}
\newcommand{\thtrep}{\theta}
\newcommand{\vpirep}{\varpi}
\newcommand{\Lamrep}{\Lambda}
\newcommand{\upsrep}{\upsilon}
\newcommand{\epsrep}{\epsilon}
\newcommand{\vphirep}{\varphi}
\newcommand{\vthtrep}{\vartheta}
\newcommand{\epsmurep}{\epsrep\murep}
\newcommand{\vectPi}{\boldsymbol{\Pirep}}
\newcommand{\vectups}{\boldsymbol{\upsrep}}
\newcommand{\vectkap}{\boldsymbol{\kaprep}}
\newcommand{\vectOme}{\boldsymbol{\Omerep}}
\newcommand{\vectLam}{\boldsymbol{\Lamrep}}

\newcommand{\tmpa}{\xi}
\newcommand{\tmpb}{\zeta}
\newcommand{\hertzian}[1]{{\Delrep}_{#1}}

\newcommand{\dragf}{d}
\newcommand{\sklife}{t_{s}}
\newcommand{\rpsi}{\psirep_1}
\newcommand{\ipsi}{\psirep_2}
\newcommand{\rome}{\omerep_1}
\newcommand{\iome}{\omerep_2}
\newcommand{\pfreq}{\omerep_o}
\newcommand{\phvel}{\vectups_{p}}
\newcommand{\grvel}{\vectups_{g}}
\newcommand{\unitkap}{\vec{\vectkap}}

\newcommand{\derivD}{\fideriv{\vectD}}

\begin{document}

\workno{GEM}{2001}{01C}
\worker{A. I. A. Adewole}
\monthdate{January 2001}
\mailto{aiaa@adequest.ca}
\work{Light Propagation For Accelerated Observers}
\shortwork{Light Propagation For Accelerated Observers}
\makefront

\begin{abstract}
We show that for an observer in translational, rotational or 
gravitational motion, a linearly polarized plane wave has two modes of propagation in a
stationary, homogeneous and isotropic medium according to Hertz's version of Maxwell's theory.
The first mode is characterized by polarization
at right angles to the direction of propagation and has a phase velocity that is controlled
by the material constants of the medium. The second mode is characterized by polarization 
along the propagation direction and has a phase velocity that is controlled by the motion
of the observer. We outline some applications of the second mode in emerging technologies.
\end{abstract}

\section{Introduction}\label{INTRO}
We have shown previously that in Hertz's version of Maxwell's theory,
an electric field $\vectD(\vectr,\scalt)$ is described by the equations
\begin{subequations}\label{EM:D0}
\begin{equation}\label{EM:D1}
\dive{\vectD} = \sigrep
\end{equation}
\begin{equation}\label{EM:D2}
\curl{\vectH} = \partial\vectD/\partial\scalt + \vectj
\end{equation}
\begin{equation}\label{EM:D3}
\vectj = \udiv{\vectD}
\end{equation}
while a magnetic field $\vectB(\vectr,\scalt)$ is described by the
equations~\footnote{Equation \eqnref{EM:B1} will not be used anywhere in this paper; 
it is included here only for reasons of completeness.}
\end{subequations}
\begin{subequations}\label{EM:B0}
\begin{equation}\label{EM:B1}
\dive{\vectB} = \delrep
\end{equation}
\begin{equation}\label{EM:B2}
\curl{\vectE} = -\partial\vectB/\partial\scalt - \vectb
\end{equation}
\begin{equation}\label{EM:B3}
\vectb = \udiv{\vectB}
\end{equation}
\end{subequations}
where $\vectu=d\vectr/dt$ is the velocity of point $\vectr$ where one is 
observing the fields at time $\scalt$ and may be loosely described as the velocity 
of an observer, detector or probe at this point~\cite{Adewole00a}. 
Our purpose 
in this paper is to use these equations to study the propagation of light relative 
to accelerated observers (or more precisely, accelerated points of observation) in a
stationary, homogeneous and isotropic medium with constitutive equations
\begin{equation}\label{EM:C}
\vectD = \epsrep\vectE,\quad\vectB = \murep\vectH
\end{equation}
where $\epsrep$ and $\murep$ are independent of both time $\scalt$ and
position $\vectr$.

We are motivated to study this problem for theoretical and practical reasons.
On the theoretical side, we need to find experimentally testable consequences
of \eqnref{EM:D3} and \eqnref{EM:B3} because these equations represent the
point of departure of Hertz's theory from Maxwell-Einstein electrodynamics of
moving bodies. Also, a rigorous classical treatment of the problem
has until now seemed to be elusive, leaving many classical physicists
to conjecture the validity of the galilean law of velocity composition for 
translating, rotating and gravitating observers (see, for example,~\cite{Hayden91}).
This conjecture can
clearly not be assumed to have unlimited validity since an electromagnetic 
wave is governed by the laws of
electrodynamics while a point particle for which the galilean law is
manifestly valid is not governed by such laws. It turns out that when the
limits of validity of this conjecture are properly studied, one finds that,
unusual as it may seem, it is possible for light to propagate with a 
velocity that depends only on the motion of the observer and not on
the permittivity or the permeability of a medium. This peculiar mode
of propagation has a number of potential applications, some of which will
be highlighted in this paper. 

Even without these novel applications, we have a practical motivation
in the fact that light propagation relative to accelerated observers can
be used to detect acceleration in devices such as accelerometers (for
translation), fiberoptic gyroscopes (for rotation), gravimeters and 
gradiometers (for gravitation). These devices are used in many major industries.
For examples, accelerometers are used extensively in the automotive industry,
fiberoptic gyroscopes are essential components of navigational systems
while gravimeters and gradiometers are indispensible tools for mineral exploration.
Hence an important goal of this work is to suggest new ways of significantly 
improving the stability and sensitivity of these devices. This goal is desirable
particularly for accelerometers because the current technologies
(capacitive, piezoelectric, piezoresistive, etc)
used in these devices are semi-mechanical in the sense that they require the
use of a proofmass whose response to acceleration is transduced according 
to Newton's second law. Although the more recent thermal technologies do not require a
proofmass, they require a temperature-sensitive material substrate and
therefore have many of the drawbacks of the semi-mechanical methods.
By providing a useable closed-form solution to the problem of light propagation
relative to accelerated observers, this work aims to provide a basis for
sensing all kinds of acceleration and acceleration gradients by 
fiberoptic methods.

We substitute \eqnref{EM:C} into \eqnref{EM:D2} and \eqnref{EM:B2} 
to get, respectively,
\begin{subequations}\label{curls}
\begin{equation}\label{curlb}
\curl{\vectB} = \murep\left(\fderiv{\vectD} + \vectj\right)
\end{equation}
\begin{equation}\label{curld}
\curl{\vectD} = -\epsrep\left(\fderiv{\vectB} + \vectb\right).
\end{equation}
\end{subequations}
We take the curl of \eqnref{curld} and use both \eqnref{vid1} and 
\eqnref{curlb} to get the electric field wave equation
\begin{equation}\label{WVEQ}
c^2[\lapl{\vectD} - \grad{(\dive{\vectD})}] = \sderiv{\vectD}{2}
 + \fderiv{\vectj} + (1/\murep)(\curl{\vectb}), 
\quad c = 1/\sqrt{\epsmurep}.
\end{equation}
Solutions of this equation will be sought as plane waves of the form
\begin{equation}\label{PLNWV}
\vectD = \vectD_0e^{i(\omega\scalt+\dprod{\vectkap}{\vectr})}
\end{equation}
where $\vectD_0, \omerep \text{ and } \vectkap$ are independent of
time \scalt\ and position \vectr. For waves of this form it is easy
to show that
\begin{subequations}\label{nabids}
\begin{equation}\label{nabid1}
\lapl{\vectD} = (\dprod{i\vectkap}{i\vectkap})\vectD = -\kaprep^2\vectD
\end{equation}
\begin{equation}\label{nabid2}
\grad{(\dive{\vectD})} = i\vectkap(\dprod{i\vectkap}{\vectD}) 
= -\vectkap(\dprod{\vectkap}{\vectD})
\end{equation}
\begin{equation}\label{nabid3}
\partial^2\vectD/\partial\scalt^2 = (\dprod{i\omerep}{i\omerep})
\vectD = -\omerep^2\vectD
\end{equation}
\begin{equation}\label{nabid4}
\begin{split}
\udiv{(\partial\vectD/\partial\scalt)}
&=\udiv{(i\omerep\vectD)}=i\omerep\udiv{\vectD}\\
&=(i\omerep)(\dprod{\vectu}{i\vectkap})\vectD
=-\omerep(\dprod{\vectu}{\vectkap})\vectD.
\end{split}
\end{equation}
\end{subequations}
We shall find it convenient to introduce the quantity
\begin{subequations}\label{abbrvs}
\begin{equation}\label{abbrv1}
\psirep = \omerep + \dprod{\vectu}{\vectkap}
\end{equation}
which is complex if $\omerep$ is complex and $\kaprep$ is real. Writing
\begin{equation}\label{abbrv2}
\psirep = \rpsi + i\ipsi,\quad 
\omerep = \rome + i\iome
\end{equation}
where $\rpsi, \ipsi, \rome, \iome$ are all real, we have from \eqnref{abbrv1}
that
\begin{equation}\label{abbrv3}
\rpsi = \rome + \dprod{\vectu}{\vectkap},
\quad \ipsi = \iome
\end{equation}
\end{subequations}
and from both \eqnref{abbrv2} and \eqnref{PLNWV} that
\begin{equation}\label{abbrv4}
\vectD = \vectD_1e^{i(\rome\scalt+\dprod{\vectkap}{\vectr})},
\quad\vectD_1(t) = \vectD_0e^{-\iome t}.
\end{equation}
We shall characterize this plane wave by its phase velocity $\phvel$, its group
velocity $\grvel$, and the time $\sklife$ required by its amplitude to decay 
by a factor of $1/e$ (called the skin time). These quantities are defined by
\begin{subequations}\label{wvprps}
\begin{equation}\label{wvprp1}
\phvel = (\rome/\kaprep)\unitkap
\end{equation}
\begin{equation}\label{wvprp2}
\grvel = \nabla_{\kaprep}\,\rome
\end{equation}
\begin{equation}\label{wvprp3}
\sklife = 1/\iome
\end{equation}
\end{subequations}
where $\unitkap$ is a unit vector in the direction of $\vectkap$.
We shall not consider the situation when $\omerep$ is real
and $\kaprep$ is complex in this paper, but the reader is encouraged 
to analyze this situation as an exercise. 

\section{Translating Observers}\label{TRANS}
We consider an observer in translational motion such that $\vectu$ is
a function of time only, i.e.,
\begin{equation}\label{trans1}
\vectu = \vectu(\scalt), \quad\vecta(\scalt) = \fideriv{\vectu}.
\end{equation}
Using \eqnref{trans1} and \eqnref{EM:D3}, we have
\begin{align}
\fideriv{\vectj}
&=\fideriv{[\udiv{\vectD}]}\nonumber\\
&=\gdiv{[\fideriv{\vectu}]}{\vectD}+\udiv{[\fideriv{\vectD}]}\nonumber\\
&= \gdiv{\vecta}{\vectD} + \udiv{(\partial\vectD/\partial\scalt)}\label{trans2}.
\end{align}
We also have
\begin{align}
\curl{\vectb} \nonumber &= \curl{\{\udiv{\vectB}\}} \beqref{EM:B3}\nonumber\\
&= \curl{\{\grad{(\dprod{\vectu}{\vectB})} - \cprod{\vectu}{(\curl{\vectB})}
- \gdiv{\vectB}{\vectu} - \cprod{\vectB}{(\curl{\vectu})} \}} \beqref{vid2}\nonumber\\
&= - \curl{\{\cprod{\vectu}{(\curl{\vectB})}\}} \beqref{vid4}\ \&\ \eqnref{trans1}\nonumber\\
&= -\{\vectu(\dive{(\curl{\vectB})})-\udiv{(\curl{\vectB})} + 
\gdiv{(\curl{\vectB})}{\vectu} - (\curl{\vectB})(\dive{\vectu}) \}\beqref{vid3}\nonumber\\
&= \udiv{(\curl{\vectB})} \beqref{vid5}\ \&\ \eqnref{trans1}\nonumber\\
&= \murep\udiv{(\partial\vectD/\partial\scalt)} + \murep\udiv{\udiv{\vectD}}\beqref{curlb}
\label{trans3}.
\end{align}
Substituting \eqnref{trans2} and \eqnref{trans3} into \eqnref{WVEQ} yields the wave equation
\begin{equation}\label{trans4}
c^2[\lapl{\vectD} - \grad{(\dive{\vectD})}] = \partial^2\vectD/\partial\scalt^2
 + 2\udiv{(\partial\vectD/\partial\scalt)} + \udiv{\udiv{\vectD}} 
+ \gdiv{\vecta}{\vectD}.
\end{equation}
For the plane wave of \eqnref{PLNWV}, it is easy to show that
\begin{subequations}\label{trans5}
\begin{align}
\gdiv{\vecta}{\vectD} 
&= (\dprod{\vecta}{i\vectkap})\vectD \nonumber\\
&= i(\dprod{\vecta}{\vectkap})\vectD
\end{align}
\begin{align}
\udiv{\udiv{\vectD}} 
&= \udiv{[(\dprod{\vectu}{i\vectkap})\vectD]} \nonumber\\
&= (\dprod{\vectu}{i\vectkap})\udiv{\vectD} \beqref{trans1}\nonumber\\
&= (\dprod{\vectu}{i\vectkap})(\dprod{\vectu}{i\vectkap})\vectD\nonumber\\
&= -(\dprod{\vectu}{\vectkap})^2\vectD.
\end{align}
\end{subequations}
Substituting \eqnref{trans5} and \eqnref{nabids} into \eqnref{trans4} leads
to the dispersion equation
\begin{equation}
c^2[-\kaprep^2\vectD + \vectkap(\dprod{\vectkap}{\vectD})]
= -\omerep^2\vectD - 2\omerep(\dprod{\vectu}{\vectkap})\vectD
- (\dprod{\vectu}{\vectkap})^2\vectD + i(\dprod{\vecta}{\vectkap})\vectD
\end{equation}
which, when simplified by taking \eqnref{abbrv1} into account, becomes
\begin{equation}\label{trans6}
-c^2\vectkap(\dprod{\vectkap}{\vectD}) =
[\psirep^2 - c^2\kaprep^2 - i(\dprod{\vecta}{\vectkap})]\vectD.
\end{equation}

\subsection{Transverse Waves ($\dprod{\vectkap}{\vectD} = 0$)}
For these waves the left hand side of \eqnref{trans6} vanishes. 
Ignoring the trivial solution $\vectD=\vectzero$, we have
\begin{equation}\label{trans7}
\psirep^2 - c^2\kaprep^2 - i(\dprod{\vecta}{\vectkap}) = 0
\end{equation}
from which it follows that $\psirep$, and therefore $\omerep$, is complex.
We then have
\begin{align}\label{trans8}
c^2\kaprep^2 + i(\dprod{\vecta}{\vectkap}) 
&= \psirep^2\nonumber\\
&= (\rpsi + i\ipsi)^2\beqref{abbrv2}\nonumber\\
&= \rpsi^2 - \ipsi^2 + 2i\rpsi\ipsi
\end{align}
which leads to the system of equations
\begin{equation}\label{trans9}
\rpsi^2 - \ipsi^2 = c^2\kaprep^2, \quad
2\rpsi\ipsi = \dprod{\vecta}{\vectkap}
\end{equation}
with solutions
\begin{subequations}\label{trans10}
\begin{equation}\label{trans10a}
\rpsi = \pm c\kaprep\dragf,\quad
\ipsi = \pm c\kaprep\sqrt{\dragf^2 - 1}
\qquad(\dprod{\vecta}{\vectkap}\ge0)
\end{equation}
\begin{equation}\label{trans10b}
\rpsi = \pm c\kaprep\dragf,\quad
\ipsi = \mp c\kaprep\sqrt{\dragf^2 - 1}
\qquad(\dprod{\vecta}{\vectkap}\le0)
\end{equation}
\end{subequations}
where all the upper signs (or all the lower signs) are to be taken
together in each set of equations, and $\dragf>0$ is given by
\begin{equation}\label{trans10x}
\dragf = \left\{\frac{1 + \sqrt{1 + \vthtrep^2}}{2} \right\}^{1/2},
\quad\vthtrep = \frac{\dprod{\vecta}{\vectkap}}{c^2\kaprep^2}.
\end{equation}
The corresponding frequency components as given by \eqnref{abbrv3} are
\begin{subequations}\label{trans11}
\begin{equation}\label{trans11a}
\rome = \pm c\kaprep\dragf - \dprod{\vectu}{\vectkap},\quad
\iome = \pm c\kaprep\sqrt{\dragf^2 - 1}
\qquad(\dprod{\vecta}{\vectkap}\ge0)
\end{equation}
\begin{equation}\label{trans11b}
\rome = \pm c\kaprep\dragf - \dprod{\vectu}{\vectkap},\quad
\iome = \mp c\kaprep\sqrt{\dragf^2 - 1}
\qquad(\dprod{\vecta}{\vectkap}\le0).
\end{equation}
\end{subequations}
Substituting these into \eqnref{wvprps} gives
\begin{subequations}\label{trans12}
\begin{equation}\label{trans12a}
\left.
\begin{aligned}
\phvel &= (\pm c\dragf - u\cos\thtrep)\unitkap\\
\grvel &= \pm(c\dragf\unitkap + \vectw) - \vectu\\
\sklife &= \pm\bigl(c\kaprep\sqrt{\dragf^2 - 1}\bigr)^{-1}
\end{aligned}
\qquad\right\}(\dprod{\vecta}{\vectkap}>0)
\end{equation}
\begin{equation}\label{trans12b}
\left.
\begin{aligned}
\phvel &= (\pm c\dragf - u\cos\thtrep)\unitkap\\
\grvel &= \pm(c\dragf\unitkap + \vectw) - \vectu\\
\sklife &= \mp\bigl(c\kaprep\sqrt{\dragf^2 - 1}\bigr)^{-1}
\end{aligned}
\qquad\right\}(\dprod{\vecta}{\vectkap}<0)
\end{equation}
\begin{equation}\label{trans12c}
\left.
\begin{aligned}
\phvel &= (\pm c - u\cos\thtrep)\unitkap\\
\grvel &= \pm c\unitkap - \vectu\\
\sklife &= \infty
\end{aligned}
\qquad\right\}(\dprod{\vecta}{\vectkap}=0)
\end{equation}
where $\thtrep$ is the angle between \vectu\ and $\vectkap$, and
\begin{equation}\label{trans12d}
\vectw = \frac{\chirep}{4c\kaprep\dragf}
  \biggl\{ \vecta - \vpirep_1\vectkap\biggr\},\quad
\chirep=\vthtrep(1+\vthtrep^2)^{-1/2},\quad
\vpirep_1=2\alprep\kaprep^{-2},\quad
\alprep=\dprod{\vecta}{\vectkap}.
\end{equation}
\end{subequations}

\subsection{Nontransverse Waves ($\dprod{\vectkap}{\vectD} \ne 0$)}
For these waves we multiply \eqnref{trans6} scalarwise by $\vectkap$ to get
\begin{equation}\label{trans14}
-c^2\kaprep^2(\dprod{\vectkap}{\vectD}) =
[\psirep^2 - c^2\kaprep^2 - i(\dprod{\vecta}{\vectkap})]
(\dprod{\vectkap}{\vectD}).
\end{equation}
Since $\dprod{\vectkap}{\vectD} \ne 0$, we obtain
\begin{equation}
\psirep^2 = i(\dprod{\vecta}{\vectkap})
\end{equation}
which leads to the system of equations
\begin{equation}
\rpsi^2 - \ipsi^2 = 0,\quad
2\rpsi\ipsi = \dprod{\vecta}{\vectkap}
\end{equation}
with solutions
\begin{subequations}
\begin{equation}
\rpsi = \pm\kaprep\etarep,\quad \ipsi = \pm\kaprep\etarep
\qquad(\dprod{\vecta}{\vectkap}\ge0)
\end{equation}
\begin{equation}
\rpsi = \pm\kaprep\etarep,\quad \ipsi = \mp\kaprep\etarep
\qquad(\dprod{\vecta}{\vectkap}\le0)
\end{equation}
\end{subequations}
where $\etarep\ge0$ satisfies
\begin{equation}
\etarep^2 = \left|\frac{\dprod{\vecta}{\vectkap}}{2\kaprep^2}\right|.
\end{equation}
The corresponding frequency components as given by \eqnref{abbrv3} are
\begin{subequations}\label{trans15}
\begin{equation}\label{trans15a}
\rome = \pm\kaprep\etarep - \dprod{\vectu}{\vectkap},\quad
\iome = \pm\kaprep\etarep
\qquad(\dprod{\vecta}{\vectkap}\ge0)
\end{equation}
\begin{equation}\label{trans15b}
\rome = \pm\kaprep\etarep - \dprod{\vectu}{\vectkap},\quad
\iome = \mp\kaprep\etarep
\qquad(\dprod{\vecta}{\vectkap}\le0).
\end{equation}
\end{subequations}
Substituting these into \eqnref{wvprps} gives
\begin{subequations}\label{trans16}
\begin{equation}\label{trans16a}
\left.
\begin{aligned}
\phvel &= (\pm\etarep - u\cos\thtrep)\unitkap\\
\grvel &= \pm\vecta/(4\kaprep\etarep) - \vectu \\
\sklife &= \pm(\kaprep\etarep)^{-1}
\end{aligned}
\qquad\right\}(\dprod{\vecta}{\vectkap}>0)
\end{equation}
\begin{equation}\label{trans16b}
\left.
\begin{aligned}
\phvel &= (\pm\etarep - u\cos\thtrep)\unitkap\\
\grvel &= \pm\vecta/(4\kaprep\etarep) - \vectu \\
\sklife &= \mp(\kaprep\etarep)^{-1}
\end{aligned}
\qquad\right\}(\dprod{\vecta}{\vectkap}<0)
\end{equation}
\begin{equation}\label{trans16c}
\left.
\begin{aligned}
\phvel &= -u\cos\thtrep\unitkap\\
\grvel &= -\vectu \\
\sklife &= \infty
\end{aligned}
\qquad\right\}(\dprod{\vecta}{\vectkap}=0)
\end{equation}
\end{subequations}
where, as before, $\thtrep$ is the angle between \vectu\ and $\vectkap$.

\section{Rotating Observers}\label{ROTAT}
We consider an observer in rotational motion with velocity
\begin{subequations}\label{rota1}
\begin{equation}\label{rota1a}
\vectu = \cprod{\vectOme}{\vectr},\quad
\vectOme = \vectOme(\scalt)
\end{equation}
and accelerations
\begin{equation}\label{rota1b}
\vectLam(\scalt) = \partial\vectOme/\partial\scalt,\quad
\vecta = \partial\vectu/\partial\scalt = \cprod{\vectLam}{\vectr}.
\end{equation}
\end{subequations}
It is easy to show from these equations that
\begin{subequations}\label{rota2}
\begin{equation}\label{rota2a}
\dive{\vectu} = 0,\quad
\curl{\vectu} = 2\vectOme
\end{equation}
and for any \vectB,
\begin{align}
\gdiv{\vectB}{\vectu}
&= \gdiv{\vectB}{(\cprod{\vectOme}{\vectr})}\beqref{rota1a}\nonumber\\
&= \cprod{(\gdiv{\vectB}{\vectOme})}{\vectr}
 + \cprod{\vectOme}{(\gdiv{\vectB}{\vectr})}\nonumber\\
&= \cprod{\vectOme}{(\gdiv{\vectB}{\vectr})}\beqref{rota1a}\nonumber\\
&= \cprod{\vectOme}{\vectB}\label{rota2b}.
\end{align}
\end{subequations}
Taking the curl of \vectb\ gives
\begin{align}\label{rota4}
\curl{\vectb} 
&= \curl{(\udiv{\vectB})} \beqref{EM:B3}\nonumber\\
&= \curl{[\grad{(\dprod{\vectu}{\vectB})} - \cprod{\vectu}{(\curl{\vectB})}
 - \gdiv{\vectB}{\vectu} - \cprod{\vectB}{(\curl{\vectu})}]} \beqref{vid2}\nonumber\\
&= \curl{(\cprod{(\curl{\vectB})}{\vectu})} - \curl{(\gdiv{\vectB}{\vectu})}
  -2\curl{(\cprod{\vectB}{\vectOme})}\beqref{vid4}\ \&\ \eqnref{rota2a}\nonumber\\
\begin{split}
&=
(\curl{\vectB})(\dive{\vectu}) - \gdiv{(\curl{\vectB})}{\vectu} +
   \udiv{(\curl{\vectB})} - \vectu(\dive{(\curl{\vectB})})\\
   &\qquad-\curl{(\gdiv{\vectB}{\vectu})}-2\curl{(\cprod{\vectB}{\vectOme})}
   \beqref{vid3}
\end{split}
\nonumber\\
\begin{split}
&=
-\gdiv{(\curl{\vectB})}{\vectu} + \udiv{(\curl{\vectB})}
   -\curl{(\gdiv{\vectB}{\vectu})}\\
&\qquad-2\curl{(\cprod{\vectB}{\vectOme})}
   \beqref{rota2a}\ \&\ \eqnref{vid5}.
\end{split}
\end{align}
Using \eqnref{rota2b} to rewrite the terms $\gdiv{(\curl{\vectB})}{\vectu}$
and $\gdiv{\vectB}{\vectu}$ respectively as $\cprod{\vectOme}{(\curl{\vectB})}$ 
and $\cprod{\vectOme}{\vectB}$, we have
\begin{align}
\curl{\vectb} 
&= \cprod{(\curl{\vectB})}{\vectOme} + \udiv{(\curl{\vectB})} +
  \curl{(\cprod{\vectOme}{\vectB})}\nonumber\\
\begin{split}
&= \cprod{(\curl{\vectB})}{\vectOme} + \udiv{(\curl{\vectB})}\\
  &\qquad+[\vectOme(\dive{\vectB}) + \gdiv{\vectB}{\vectOme} - \vectB(\dive{\vectOme})
  - \gdiv{\vectOme}{\vectB}]\beqref{vid3}
\end{split}
\nonumber\\ 
&= \cprod{(\curl{\vectB})}{\vectOme} + \udiv{(\curl{\vectB})}
  +\vectOme(\dive{\vectB})-\gdiv{\vectOme}{\vectB}\beqref{rota1a}\label{rota5}.
\end{align}
This equation is valid for any \vectB\ provided \vectu\ is given by \eqnref{rota1}.
In particular, replacing \vectB\ with \vectD\ gives
\begin{subequations}\label{rota6}
\begin{equation}\label{rota6a}
\curl{\vectj} = \cprod{(\curl{\vectD})}{\vectOme} + \udiv{(\curl{\vectD})}
  +\vectOme(\dive{\vectD})-\gdiv{\vectOme}{\vectD}
\end{equation}
while replacing \vectB\ with $\curl{\vectB}$ gives
\begin{align}
\begin{split}
\curl{(\udiv{(\curl{\vectB})})} 
&=\cprod{(\curl{(\curl{\vectB})})}{\vectOme} + \udiv{(\curl{(\curl{\vectB})})}\\
  &\qquad+\vectOme(\dive{(\curl{\vectB})})-\gdiv{\vectOme}{(\curl{\vectB})}
\end{split}
\nonumber\\
\begin{split}
&= -\cprod{\vectOme}{(\curl{(\curl{\vectB})})} + \udiv{(\curl{(\curl{\vectB})})}\\
  &\qquad-\gdiv{\vectOme}{(\curl{\vectB})}\beqref{vid5}.
\end{split}
\label{rota6b}
\end{align}
\end{subequations}
Our goal is to obtain an expression from $\curl{\vectb}$ in terms of
$\curl{\vectB}$ so that \vectB\ can be eliminated with \eqnref{curlb}.
For this reason we need to eliminate the last two terms of \eqnref{rota5}.
Taking the curl of both sides of this equation, 
\begin{align}
\begin{split}
\curl{(\curl{\vectb})}
&=\curl{[\cprod{(\curl{\vectB})}{\vectOme}]} + \curl{[\udiv{(\curl{\vectB})}]}\\
  &\qquad\curl{[\vectOme(\dive{\vectB})]} - \curl{[\gdiv{\vectOme}{\vectB}]}
\end{split}
\nonumber\\
\begin{split}
&=\curl{[\cprod{(\curl{\vectB})}{\vectOme}]} + \curl{[\udiv{(\curl{\vectB})}]}\\
  &\qquad-\cprod{\vectOme}{(\grad{(\dive{\vectB})})} - \gdiv{\vectOme}{(\curl{\vectB})}
\beqref{vid6}\ \&\ \eqnref{rota1a}
\end{split}
\nonumber\\
\begin{split}
&=(\curl{\vectB})(\dive{\vectOme}) - \gdiv{(\curl{\vectB})}{\vectOme}
  +\gdiv{\vectOme}{(\curl{\vectB})} - \vectOme(\dive{(\curl{\vectB})})\\
  &\qquad+ \curl{(\udiv{(\curl{\vectB})})}-\cprod{\vectOme}{(\grad{(\dive{\vectB})})} 
     - \gdiv{\vectOme}{(\curl{\vectB})}\beqref{vid3}
\end{split}
\nonumber\\
&=\curl{(\udiv{(\curl{\vectB})})}-\cprod{\vectOme}{(\grad{(\dive{\vectB})})} 
\beqref{rota1a}\ \&\ \eqnref{vid5}\nonumber\\
\begin{split}
&= -\cprod{\vectOme}{(\curl{(\curl{\vectB})})} + \udiv{(\curl{(\curl{\vectB})})}\\
  &\qquad-\gdiv{\vectOme}{(\curl{\vectB})}-\cprod{\vectOme}{(\grad{(\dive{\vectB})})}
   \beqref{rota6b}.
\end{split}
\label{rota7}
\end{align}
Multiplying scalarwise by $\vectOme$ gives an expression
\begin{equation}\label{rota8}
\dprod{\vectOme}{[\curl{(\curl{\vectb})}]}=
\dprod{\vectOme}{\{[\udiv{(\curl{(\curl{\vectB})})}]
  -[\gdiv{\vectOme}{(\curl{\vectB})}]\}}
\end{equation}
from which \vectB\ can be eliminated.
To eliminate \vectB, we use \eqnref{curlb} and \eqnref{EM:D3} to get
\begin{align}\label{rota9}
\gdiv{\vectOme}{(\curl{\vectB})} 
&= \murep\gdiv{\vectOme}{[\derivD + \udiv{\vectD}]}.
\end{align}
Similarly, by writing $\vectPi = \curl{\vectD}$ and using \eqnref{curlb},
\begin{align}\label{rota10}
\udiv{(\curl{(\curl{\vectB})})}
&=\murep\udiv{(\fideriv{\vectPi})} + \murep\udiv{(\curl{\vectj})}\nonumber\\
\begin{split}
&=\murep\udiv{[\fideriv{\vectPi}+\udiv{\vectPi}-\gdiv{\vectOme}{\vectD}]}\\
&\qquad+\murep\udiv{(\cprod{\vectPi}{\vectOme})}+\murep\udiv{
  [\vectOme(\dive{\vectD})]}\beqref{rota6a}
\end{split}
\nonumber\\
\begin{split}
&=\murep\udiv{[\fideriv{\vectPi}+\udiv{\vectPi}-\gdiv{\vectOme}{\vectD}]}\\
  &\qquad+\murep[\cprod{(\udiv{\vectPi})}{\vectOme}+\cprod{\vectPi}{(\udiv{\vectOme})}]\\
  &\qquad+\murep[(\udiv{\vectOme})(\dive{\vectD})+\vectOme\udiv{(\dive{\vectD})}]
    \beqref{vid7}\ \&\ \eqnref{vid8}
\end{split}
\nonumber\\
\begin{split}
&=\murep\udiv{[\fideriv{\vectPi}+\udiv{\vectPi}-\gdiv{\vectOme}{\vectD}]}\\
  &\qquad-\murep\cprod{\vectOme}{(\udiv{\vectPi})}+\murep\vectOme
   (\dprod{\vectu}{\grad{(\dive{\vectD})}})\beqref{rota1a}.
\end{split}
\end{align}
It follows from \eqnref{rota8}, \eqnref{rota9}, \eqnref{rota10} and
\eqnref{EM:D1} that
\begin{subequations}
\begin{align}\label{rota11}
\dprod{\vectOme}{[\curl{(\curl{\vectb})}]}
&=\murep\dprod{\vectOme}{\{\vectOme(\dprod{\vectu}{\grad{\sigrep}})
  +\udiv{[\hertzian{\scalu}\vectPi-\gdiv{\vectOme}{\vectD}]}-\gdiv{\vectOme}{\hertzian{\scalu}\vectD}\}}
\end{align}
where we have introduced the convective or hertzian operator
\begin{equation}\label{rota12}
\hertzian{\scalu}\equiv \fderiv{} + \udiv{}
\end{equation}
\end{subequations}
for convenience.

\subsection{The Wave Equation}
From \eqnref{WVEQ} and \eqnref{EM:D1}, we have
\begin{equation}\label{rota13}
(1/\murep)(\curl{\vectb})=
c^2(\lapl{\vectD} - \grad{\sigrep})-\sideriv{\vectD}{2}-\fideriv{\vectj}. 
\end{equation}
Let us take the curl of this equation to get
\begin{align}
(1/\murep)(\curl{(\curl{\vectb})})
&=c^2\curl{(\lapl{\vectD}-\grad{\sigrep})}-\sideriv{\vectPi}{2}
  -\fideriv{(\curl{\vectj})}\nonumber\\
&=c^2\curl{(\lapl{\vectD})}-\sideriv{\vectPi}{2}-\fderiv{}\biggl[\cprod{\vectPi}
  {\vectOme}+\udiv{\vectPi}\nonumber\\
   &\qquad+\sigrep\vectOme-\gdiv{\vectOme}{\vectD}\biggr]\beqref{rota6a}\ \&\ \eqnref{vid4}
\nonumber\\
\begin{split}
&=c^2\curl{(\lapl{\vectD})}-\sideriv{\vectPi}{2}+\cprod{\vectOme}{(\fideriv{\vectPi})}
   +\cprod{\vectLam}{\vectPi}-\gdiv{\vecta}{\vectPi}\\
 &\qquad-\udiv{(\fideriv{\vectPi})}-\sigma\vectLam-\vectOme(\fideriv{\sigrep})
  +\gdiv{\vectLam}{\vectD}\\
 &\qquad+\gdiv{\vectOme}{(\fideriv{\vectD})}\beqref{rota1b}.
\end{split}
\label{rota14}
\end{align}
From \eqnref{rota11} and \eqnref{rota14}, we finally obtain the wave equation
\begin{align}\label{rota15}
\begin{split}
c^2\dprod{\vectOme}{[\curl{(\lapl{\vectD})}]}
&=\dprod{\vectOme}{}\biggl[\sderiv{\vectPi}{2}+\sigrep\vectLam+\vectOme\fderiv{\sigrep}
  -\cprod{\vectLam}{\vectPi}+\gdiv{\vecta}{\vectPi}+2\udiv{\fderiv{\vectPi}}\\
 &\quad-\gdiv{\vectLam}{\vectD}-2\gdiv{\vectOme}{\fderiv{\vectD}}+\vectOme(
   \dprod{\vectu}{\grad{\sigrep}})+\udiv{(\udiv{\vectPi})}\\
 &\quad-\udiv{(\gdiv{\vectOme}{\vectD})}-\gdiv{\vectOme}{(\udiv{\vectD})}\biggr].
\end{split}
\end{align}
The various terms of this equation can be evaluated
for the plane wave of \eqnref{PLNWV} as follow. Introducing
\begin{subequations}\label{rota16}
\begin{equation}\label{rota16a}
\betrep = \dprod{\vectOme}{(\cprod{\vectkap}{\vectD})},
\end{equation}
we have
\begin{align}\label{rota16b}
\dprod{\vectOme}{[\curl{(\lapl{\vectD})}]}
&=\dprod{\vectOme}{[\curl{((i\vectkap)^2\vectD)}]}\nonumber\\
&=-\kaprep^2\dprod{\vectOme}{[\curl{\vectD}]}\nonumber\\
&=-\kaprep^2\dprod{\vectOme}{[\cprod{i\vectkap}{\vectD}]}\nonumber\\
&=-i\kaprep^2\betrep
\end{align}
\begin{align}\label{rota16c}
\dprod{\vectOme}{\biggl[\sderiv{\vectPi}{2}\biggr]}
&=\dprod{\vectOme}{\biggl[\sderiv{}{2}(\curl{\vectD})\biggr]}\nonumber\\
&=\dprod{\vectOme}{\biggl[\sderiv{}{2}(\cprod{i\vectkap}{\vectD})\biggr]}\nonumber\\
&=\dprod{\vectOme}{\biggl[\cprod{i\vectkap}{\sderiv{\vectD}{2}}\biggr]}\nonumber\\
&=\dprod{\vectOme}{[\cprod{i\vectkap}{(i\omerep)^2\vectD}]}\nonumber\\
&=-i\omerep^2\betrep
\end{align}
\begin{align}\label{rota16d}
\dprod{\vectOme}{[\sigrep\vectLam]}
&=\dprod{\vectOme}{[\vectLam(\dive{\vectD})]}\nonumber\\
&=\dprod{\vectOme}{[\vectLam(\dprod{i\vectkap}{\vectD})]}\nonumber\\
&=i(\dprod{\vectkap}{\vectD})(\dprod{\vectOme}{\vectLam})
\end{align}
\begin{align}\label{rota16e}
\dprod{\vectOme}{\biggl[\vectOme\fderiv{\sigrep}\biggr]}
&=\Omerep^2\fderiv{}(\dive{\vectD})\nonumber\\
&=\Omerep^2\fderiv{}(\dprod{i\vectkap}{\vectD})\nonumber\\
&=i\Omerep^2\dprod{\vectkap}{\fderiv{\vectD}}\nonumber\\
&=i\Omerep^2\dprod{\vectkap}{(i\omerep\vectD)}\nonumber\\
&=-\omerep\Omerep^2(\dprod{\vectkap}{\vectD})
\end{align}
\begin{align}\label{rota16f}
\dprod{\vectOme}{[\cprod{\vectLam}{\vectPi}]}
&=\dprod{\vectOme}{[\cprod{\vectLam}{(\curl{\vectD})}]}\nonumber\\
&=\dprod{\vectOme}{[\cprod{\vectLam}{(\cprod{i\vectkap}{\vectD})}]}\nonumber\\
&=-i\dprod{\vectLam}{[\cprod{\vectOme}{(\cprod{\vectkap}{\vectD})}]}
\end{align}
\begin{align}\label{rota16g}
\dprod{\vectOme}{[\gdiv{\vecta}{\vectPi}]}
&=\dprod{\vectOme}{[\gdiv{\vecta}{(\curl{\vectD})}]}\nonumber\\
&=\dprod{\vectOme}{[\gdiv{\vecta}{(\cprod{i\vectkap}{\vectD})}]}\nonumber\\
&=i\dprod{\vectOme}{[\cprod{\vectkap}{(\gdiv{\vecta}{\vectD})}]}\nonumber\\
&=i\dprod{\vectOme}{[\cprod{\vectkap}{((\dprod{\vecta}{i\vectkap})\vectD)}]}\nonumber\\
&=-\dprod{\vectOme}{[(\dprod{\vecta}{\vectkap})(\cprod{\vectkap}{\vectD})]}\nonumber\\
&=-(\dprod{\vecta}{\vectkap})\betrep
\end{align}
\begin{align}\label{rota16h}
\dprod{\vectOme}{\biggl[\udiv{\fderiv{\vectPi}}\biggr]}
&=\dprod{\vectOme}{\biggl[\udiv{\fderiv{}(\curl{\vectD})}\biggr]}\nonumber\\
&=\dprod{\vectOme}{\biggl[\udiv{\fderiv{}(\cprod{i\vectkap}{\vectD})}\biggr]}\nonumber\\
&=i\dprod{\vectOme}{\biggl[\udiv{\biggl(\cprod{\vectkap}{\fderiv{\vectD}}\biggr)}\biggr]}\nonumber\\
&=i\dprod{\vectOme}{[\udiv{(\cprod{\vectkap}{i\omerep\vectD})}]}\nonumber\\
&=-\omerep\dprod{\vectOme}{[\cprod{\vectkap}{(\udiv{\vectD})}]}\nonumber\\
&=-\omerep\dprod{\vectOme}{[\cprod{\vectkap}{((\dprod{\vectu}{i\vectkap})\vectD)}]}\nonumber\\
&=-i\omerep(\dprod{\vectu}{\vectkap})\betrep
\end{align}
\begin{align}\label{rota16i}
\dprod{\vectOme}{[\gdiv{\vectLam}{\vectD}]}
&=\dprod{\vectOme}{[(\dprod{\vectLam}{i\vectkap})\vectD]}\nonumber\\
&=i(\dprod{\vectLam}{\vectkap})(\dprod{\vectOme}{\vectD})
\end{align}
\begin{align}\label{rota16j}
\dprod{\vectOme}{\biggl[\gdiv{\vectOme}{\fderiv{\vectD}}\biggr]}
&=\dprod{\vectOme}{[\gdiv{\vectOme}{(i\omerep\vectD)}]}\nonumber\\
&=i\omerep\dprod{\vectOme}{[\gdiv{\vectOme}{\vectD}]}\nonumber\\
&=i\omerep\dprod{\vectOme}{[(\dprod{\vectOme}{i\vectkap})\vectD]}\nonumber\\
&=-\omerep(\dprod{\vectOme}{\vectD})(\dprod{\vectOme}{\vectkap})
\end{align}
\begin{align}\label{rota16k}
\dprod{\vectOme}{[\vectOme(\dprod{\vectu}{\grad{\sigrep}})]}
&=\Omerep^2[\dprod{\vectu}{\grad{(\dive{\vectD})}}]\nonumber\\
&=\Omerep^2[\dprod{\vectu}{\grad{(\dprod{i\vectkap}{\vectD})}}]\nonumber\\
&=i\Omerep^2\dprod{\vectu}{[\grad{(\dprod{\vectkap}{\vectD})}]}\nonumber\\
&=i\Omerep^2\dprod{\vectu}{[\cprod{\vectkap}{(\curl{\vectD})}
  +\gdiv{\vectkap}{\vectD}+\gdiv{\vectD}{\vectkap}+\cprod{\vectD}{(\curl{\vectkap})}]}\nonumber\\
&=i\Omerep^2\dprod{\vectu}{[\cprod{\vectkap}{(\cprod{i\vectkap}{\vectD})}
  +(\dprod{\vectkap}{i\vectkap})\vectD]}\nonumber\\
&=-\Omerep^2\dprod{\vectu}{[\cprod{\vectkap}{(\cprod{\vectkap}{\vectD})}
  +\kaprep^2\vectD]}\nonumber\\
&=-\Omerep^2(\dprod{\vectu}{\vectkap})(\dprod{\vectkap}{\vectD})
\end{align}
\begin{align}\label{rota16l}
\dprod{\vectOme}{[\udiv{(\udiv{\vectPi})}]}
&=\dprod{\vectOme}{[\udiv{(\udiv{(\curl{\vectD})})}]}\nonumber\\
&=\dprod{\vectOme}{[\udiv{(\udiv{(\cprod{i\vectkap}{\vectD})})}]}\nonumber\\
&=i\dprod{\vectOme}{[\udiv{(\cprod{\vectkap}{(\udiv{\vectD})})}]}\nonumber\\
&=i\dprod{\vectOme}{[\udiv{(\cprod{\vectkap}{((\dprod{\vectu}{i\vectkap})\vectD)})}]}\nonumber\\
&=-\dprod{\vectOme}{[\udiv{((\dprod{\vectu}{\vectkap})(\cprod{\vectkap}{\vectD}))}]}\nonumber\\
&=-\dprod{\vectOme}{[(\cprod{\vectkap}{\vectD})\{\udiv{(\dprod{\vectu}{\vectkap})}\}+
  (\dprod{\vectu}{\vectkap})\{\udiv{(\cprod{\vectkap}{\vectD})}\}]}\nonumber\\
&=-\betrep\dprod{\vectu}{[\grad{(\dprod{\vectu}{\vectkap})}]}-(\dprod{\vectu}{\vectkap})
  \dprod{\vectOme}{[\udiv{(\cprod{\vectkap}{\vectD})}]}\nonumber\\
\begin{split}
&=-\betrep\dprod{\vectu}{[\cprod{\vectu}{(\curl{\vectkap})}+\udiv{\vectkap}
  +\gdiv{\vectkap}{\vectu}+\cprod{\vectkap}{(\curl{\vectu})}]}\\
  &\qquad-(\dprod{\vectu}{\vectkap})\dprod{\vectOme}{[\cprod{\vectkap}{\udiv{\vectD}}]}\nonumber\\
\end{split}
\nonumber\\
&=-\betrep\dprod{\vectu}{[\cprod{\vectOme}{\vectkap}+2\cprod{\vectkap}{\vectOme}]}
  -(\dprod{\vectu}{\vectkap})\dprod{\vectOme}{[\cprod{\vectkap}{(\dprod{\vectu}
   {i\vectkap})\vectD}]}\beqref{rota2b}\nonumber\\
&=-\betrep[\dprod{\vectu}{(\cprod{\vectkap}{\vectOme})}+i(\dprod{\vectu}{\vectkap})^2]
\end{align}
\begin{align}\label{rota16m}
\dprod{\vectOme}{[\udiv{\{\gdiv{\vectOme}{\vectD}\}}]}
&=\dprod{\vectOme}{[\udiv{ \{(\dprod{\vectOme}{i\vectkap})\vectD\} }]}\nonumber\\
&=i(\dprod{\vectOme}{\vectkap})\dprod{\vectOme}{[\udiv{\vectD}]}\nonumber\\
&=i(\dprod{\vectOme}{\vectkap})\dprod{\vectOme}{[(\dprod{\vectu}{i\vectkap})\vectD]}\nonumber\\
&=-(\dprod{\vectOme}{\vectkap})(\dprod{\vectu}{\vectkap})(\dprod{\vectOme}{\vectD})
\end{align}
\begin{align}\label{rota16n}
\dprod{\vectOme}{[\gdiv{\vectOme}{\{\udiv{\vectD}\}}]}
&=\dprod{\vectOme}{[\gdiv{\vectOme}{\{(\dprod{\vectu}{i\vectkap})\vectD\}}]}\nonumber\\
&=i\dprod{\vectOme}{[(\dprod{\vectu}{\vectkap})\{\gdiv{\vectOme}{\vectD}\}
  +\vectD\{\dprod{\vectOme}{\grad{(\dprod{\vectu}{\vectkap})}}\}]}\nonumber\\
\begin{split}
&=i\dprod{\vectOme}{}[(\dprod{\vectu}{\vectkap})\{(\dprod{\vectOme}{i\vectkap})\vectD\}
  +\vectD\{\dprod{\vectOme}{}(\cprod{\vectu}{(\curl{\vectkap})}+\udiv{\vectkap}\\
&\qquad+\gdiv{\vectkap}{\vectu}+\cprod{\vectkap}{(\curl{\vectu})})\}]
\end{split}
\nonumber\\
&=-(\dprod{\vectOme}{\vectkap})(\dprod{\vectu}{\vectkap})(\dprod{\vectOme}{\vectD})
 +i(\dprod{\vectOme}{\vectD})\dprod{\vectOme}{[\cprod{\vectOme}{\vectkap}+
  2\cprod{\vectkap}{\vectOme}]}\beqref{rota2b}\nonumber\\
&=-(\dprod{\vectOme}{\vectkap})(\dprod{\vectu}{\vectkap})(\dprod{\vectOme}{\vectD}).
\end{align}
\end{subequations}
Substituting \eqnref{rota16} into \eqnref{rota15} leads to the dispersion equation
\begin{subequations}\label{rota17}
\begin{equation}\label{rota17a}
\betrep[\alprep+i(\psirep^2-c^2\kaprep^2)]=\psirep\tmpa+i\tmpb
\end{equation}
where
\begin{equation}\label{rota17b}
\begin{aligned}
\alprep &=\dprod{\vectkap}{(\vecta+\cprod{\vectOme}{\vectu})}\\
\tmpa   &=2(\dprod{\vectOme}{\vectD})(\dprod{\vectOme}{\vectkap})
  -\Omerep^2(\dprod{\vectkap}{\vectD})\\
\tmpb   &=(\dprod{\vectkap}{\vectD})(\dprod{\vectOme}{\vectLam})
  -(\dprod{\vectLam}{\vectD})(\dprod{\vectOme}{\vectkap}).
\end{aligned}
\end{equation}
\end{subequations}

\subsection{Coplanar Waves ($\betrep=0$)}
For these waves the left hand side of \eqnref{rota17a} vanishes and
on using \eqnref{abbrv2}, we have
\begin{equation}\label{rota18}
\tmpa\rpsi+i(\tmpb+\tmpa\ipsi)=0
\end{equation}
which implies that
\begin{subequations}\label{rota19}
\begin{equation}\label{rota19a}
\tmpa\rpsi=0
\end{equation}
\begin{equation}\label{rota19b}
\tmpb+\tmpa\ipsi=0.
\end{equation}
\end{subequations}
If $\tmpa=0$ by \eqnref{rota19a}, then $\tmpb=0$ by \eqnref{rota19b} and since
$\betrep=0$ for coplanar waves, the dispersion equations \eqnref{rota17a} and 
\eqnref{rota18} are trivially satisfied. To
obtain a nontrivial solution, we shall require that $\tmpa\ne0$. With this condition,
we have from \eqnref{rota19} that
\begin{equation}\label{rota20}
\rpsi=0,\quad
\ipsi=-\tmpb/\tmpa.
\end{equation}
The corresponding frequency components are given by \eqnref{abbrv3} as
\begin{equation}\label{rota21}
\rome = -\dprod{\vectu}{\vectkap},\quad
\iome =-\tmpb/\tmpa.
\end{equation}
Substituting these into \eqnref{wvprps} gives
\begin{equation}\label{rota22}
\begin{aligned}
\phvel &=-u\cos\theta\unitkap\\
\grvel &=-\vectu\\
\sklife &=-\tmpa/\tmpb.
\end{aligned}
\end{equation}
Equation \eqnref{rota22} shows that when $\tmpa$ and $\tmpb$ have the same
signs, the wave grows while if they have opposite signs, the wave attenuates.
For rotation with a time independent angular velocity $\vectOme$, we have
$\vectLam=\vectzero$, whence $\tmpb=0$ and the wave does not grow or attenuate.

\subsection{Noncoplanar Waves ($\betrep\ne0$)}
For these waves we divide both sides of \eqnref{rota17a} by $\betrep$ 
and multiply by $i$ to get
\begin{equation}\label{rota23}
\psirep^2-c^2\kaprep^2-(\tmpb/\betrep)-i(\alprep-(\tmpa/\betrep)\psirep) = 0
\end{equation}
which, on using \eqnref{abbrv2}, leads to the system of equations
\begin{subequations}\label{rota24}
\begin{equation}\label{rota24a}
\rpsi^2-\ipsi^2-(\tmpa/\betrep)\ipsi-c^2\kaprep^2-(\tmpb/\betrep)=0
\end{equation}
\begin{equation}\label{rota24b}
(\tmpa/\betrep)\rpsi+2\rpsi\ipsi-\alprep=0
\end{equation}
\end{subequations}
with solutions
\begin{subequations}\label{rota25}
\begin{equation}\label{rota25a}
\rpsi=\pm c\kaprep\dragf,\quad
\ipsi=\frac{1}{2}\left(\pm\frac{\alprep}{c\kaprep\dragf}-\frac{\tmpa}{\betrep}\right)
\end{equation}
where $\dragf\ge0$ is given by
\begin{equation}\label{rota25b}
\dragf=\gamrep\left\{\frac{1+\sqrt{1+\vthtrep^2}}{2}\right\}^{1/2},\quad
\gamrep^2=\left|1+\frac{\betrep\tmpb-\tmpa^2}{c^2\kaprep^2\betrep^2}\right|,\quad
\vthtrep=\frac{\alprep}{c^2\kaprep^2\gamrep^2}.
\end{equation}
\end{subequations}
The corresponding frequency components are given by \eqnref{abbrv3} as
\begin{equation}\label{rota26}
\rome=\pm c\kaprep\dragf-\dprod{\vectu}{\vectkap},\quad
\iome=\frac{1}{2}\left(\pm\frac{\alprep}{c\kaprep\dragf}-\frac{\tmpa}{\betrep}\right).
\end{equation}
Substituting these into \eqnref{wvprps} gives
\begin{subequations}\label{rota28}
\begin{equation}\label{rota28a}
\begin{aligned}
\phvel &=(\pm c\dragf-u\cos\theta)\unitkap\\
\grvel &=\pm(c\dragf\unitkap + \vectw) - \vectu\\
\sklife &=2c\kaprep\betrep\dragf/(\pm\alprep\betrep-c\kaprep\dragf\tmpa)
\end{aligned}
\end{equation}
where
\begin{equation}\label{rota28b}
\vectw=\frac{\chirep}{4c\kaprep\dragf}
\biggl\{\vecta +\cprod{\vectOme}{\vectu}-\vpirep_1\vectkap\biggr\}
+\vpirep_2(\cprod{\vectD}{\vectOme})+\vpirep_3\vectOme-\vpirep_4\vectD 
\end{equation}
and
\begin{equation}\label{rota28c}
\begin{split}
&\vpirep_1=\frac{2}{\kaprep^2}\left\{
  \alprep-\frac{4c\kaprep\dragf\betrep\vpirep_2}{\chirep}\right\},\quad
\vpirep_2=\biggl\{\tmpb-\frac{\tmpa^2}{\betrep}\biggr\}\Gamrep\\
&\vpirep_3=\biggl\{\betrep(\dprod{\vectLam}{\vectD})
  +4\tmpa(\dprod{\vectOme}{\vectD})\biggr\}\Gamrep,\quad
\vpirep_4=\biggl\{2\tmpa\Omerep^2+\betrep(\dprod{\vectOme}{\vectLam})\biggr\}\Gamrep\\
&\chirep = \vthtrep(1+\vthtrep^2)^{-1/2},\quad
\Gamrep=\frac{\alprep\chirep-2c^2\kaprep^2\dragf^2}
  {4\dragf\gamrep^2c^3\kaprep^3\betrep^2}.
\end{split}
\end{equation}
\end{subequations}
Equation \eqnref{rota28a} shows that a wave propagating in the positive direction
(i.e., corresponding to the positive signs) with
$\alprep\ge0, \betrep\ge0$ and $\tmpa\ge0$ will grow when $\alprep\betrep<c\kaprep\dragf\tmpa$,
attenuate when $\alprep\betrep>c\kaprep\dragf\tmpa$ and neither attenuate nor grow
when $\alprep\betrep=c\kaprep\dragf\tmpa$. Moreover, if
\begin{equation}\label{rota29}
\tmpa=c\kaprep\betrep\left(1+\frac{\tmpb}{c^2\kaprep^2\betrep}\right)^{1/2}
\end{equation}
then $\gamrep=0$ and $\dragf=0$ by \eqnref{rota25b}. It follows~\footnote{We remark here
that the presence of $\gamrep$ in the denominator of the expression for $\vthtrep$ in 
\eqnref{rota25b} does not involve a division by zero. To see this, it is sufficient to
rewrite the expression for $\dragf$ in \eqnref{rota25b} by putting $\gamrep$ back 
into the radicand --- it was factored out of the radicand before 
\eqnref{rota29} was contemplated.} 
from \eqnref{rota28a} that
under this condition, the phase velocity of the wave will become independent of $c$ 
and will depend only on the projection of \vectu\ on $\vectkap$. 
By \eqnref{rota17b} and \eqnref{rota29}, this peculiar 
situation arises for an observer rotating with a time independent angular velocity 
when $\tmpa=c\kaprep\betrep$. For waves polarized at right angles to the wave 
vector ($\dprod{\vectkap}{\vectD}=0$), this condition can be written as
\begin{equation}\label{rota30}
\pfreq=2f\Omerep,\quad
\pfreq=c\kaprep,\quad
f=\cos\thtrep_1\cos\thtrep_2\sec\thtrep_3
\end{equation}
where $\thtrep_1,\thtrep_2,\thtrep_3$ are respectively the direction angles of
$\vectOme$ in a righthanded cartesian coordinate system formed by \vectD, 
$\vectkap$ and $\cprod{\vectkap}{\vectD}$. It should be noted that if
$\dragf=0$ holds with exactitude, then $\sklife=0$ holds exactly by 
\eqnref{rota28a}, which means that the wave will attenuate completely
before it has a chance to travel a sensible distance. It follows that, in practice,
one should expect $\dragf=0$ to hold only approximately.

\section{Gravitating Observers}\label{GRAVI}
We consider an observer in gravitational motion with acceleration
\begin{equation}\label{grav1}
\vectg=-\scalG\vectr/\scalr^3
\end{equation}
where \scalG\ is independent of time \scalt\ and position \vectr. The velocity
\vectu\ of the observer satisfies the familiar equations~(\cite{Heinbockel96}, pg 152)
\begin{subequations}\label{grav2}
\begin{equation}\label{grav2a}
\cprod{\vecth}{\vectu}=\vectz+\scalG\vectr/\scalr
\end{equation}
\begin{equation}\label{grav2b}
\cprod{\vectu}{\vectr}=\vecth
\end{equation}
\end{subequations}
where \vectz\ and \vecth\ are independent of time $\scalt$ and position $\vectr$. By taking
the vector product of \eqnref{grav2a} with \vecth\ and noticing that
$\dprod{\vectu}{\vecth}=0$ by \eqnref{grav2b}, we get
\begin{equation}\label{grav3}
\vectu=(\cprod{\vectz}{\vecth})/ \scalh^2+
(\cprod{\scalG\vectr}{\vecth})/ \scalr\scalh^2
\end{equation}
from which we easily obtain
\begin{subequations}\label{grav4}
\begin{equation}\label{grav4a}
\dive{\vectu} = 0,\quad \curl{\vectu}=-\scalG\vecth/\scalr\scalh^2
\end{equation}
\begin{equation}\label{grav4b}
\dive{\vectOme}=0,\quad
\curl{\vectOme}=(\cprod{\vectOme}{\vectr})/\scalr^2
=(\cprod{\vecth}{\vectg})/2\scalh^2
\end{equation}
where, for convenience, we have introduced the angular velocity
\begin{equation}\label{grav4c}
\vectOme=(\curl{\vectu})/2=-\scalG\vecth/2\scalr\scalh^2
=\scalG(\cprod{\vectr}{\vectu})/2\scalr\scalh^2.
\end{equation}
\end{subequations}
Since \vectu\ does not depend explicitly on time by \eqnref{grav3}, we have
\begin{align}\label{grav5}
\fideriv{\vectj}
&=\fideriv{[\udiv{\vectD}]}\nonumber\\
&=\gdiv{[\fideriv{\vectu}]}{\vectD}+\udiv{[\fideriv{\vectD}]}\nonumber\\
&=\udiv{(\fideriv{\vectD})}.
\end{align}
It is easy to see that \eqnref{rota4} holds for any \vectu\ and any \vectB\
provided $\vectOme=(\curl{\vectu})/2$. Let us rewrite
this equation here for convenience, i.e.,
\begin{equation}\label{grav6}
\curl{\vectb}
=-\gdiv{(\curl{\vectB})}{\vectu} + \udiv{(\curl{\vectB})}
   -\curl{(\gdiv{\vectB}{\vectu})}-2\curl{(\cprod{\vectB}{\vectOme})}.
\end{equation}
Instead of \eqnref{rota2b}, we now have
\begin{align}\label{grav7}
\gdiv{\vectB}{\vectu}
&=\gdiv{\vectB}{}\biggl\{\frac{\cprod{\vectz}{\vecth}}{\scalh^2}
  +\frac{\cprod{\scalG\vectr}{\vecth}}{\scalr\scalh^2}\biggr\}\beqref{grav3}\nonumber\\
&=\gdiv{\vectB}{}\biggl\{\frac{\cprod{\scalG\vectr}{\vecth}}{\scalr\scalh^2}\biggr\}\nonumber\\
&=\frac{\scalG}{\scalh^2}\biggl\{\{\gdiv{\vectB}{(1/\scalr)}\}(\cprod{\vectr}{\vecth})
  +(1/\scalr)\{\gdiv{\vectB}{(\cprod{\vectr}{\vecth})}\}\biggr\}\nonumber\\
&=\frac{\scalG}{\scalh^2}\biggl\{(\cprod{\vectr}{\vecth})\{\dprod{\vectB}{\grad{(1/\scalr)}}\}
  +(1/\scalr)\{\cprod{[\gdiv{\vectB}{\vectr}]}{\vecth}+\cprod{\vectr}{[\gdiv{\vectB}{\vecth}]}\}
  \biggr\}\nonumber\\
&=\frac{\scalG}{\scalh^2}\biggl\{-\frac{(\dprod{\vectB}{\vectr})(\cprod{\vectr}{\vecth})}{\scalr^3}
  +\frac{\cprod{\vectB}{\vecth}}{\scalr}\biggr\}\nonumber\\
&=2\biggl\{\frac{(\dprod{\vectB}{\vectr})(\cprod{\vectr}{\vectOme})}{\scalr^2}
  -\cprod{\vectB}{\vectOme}\biggr\}\beqref{grav4c}.
\end{align}
In the same way we get the following equations
\begin{subequations}\label{grav8}
\begin{equation}\label{grav8a}
\gdiv{\vectkap}{\vectu}
=2\biggl\{\frac{(\dprod{\vectkap}{\vectr})(\cprod{\vectr}{\vectOme})}{\scalr^2}
  -\cprod{\vectkap}{\vectOme}\biggr\}
\end{equation}
\begin{equation}\label{grav8c}
\gdiv{(\curl{\vectB})}{\vectu}
=2\biggl\{\frac{(\dprod{(\curl{\vectB})}{\vectr})(\cprod{\vectr}{\vectOme})}{\scalr^2}
  -\cprod{(\curl{\vectB})}{\vectOme}\biggr\}
\end{equation}
and
\begin{align}\label{grav8b}
\gdiv{\vectr}{\vectOme}
&=\gdiv{\vectr}{(-\scalG\vecth/2\scalr\scalh^2)}\beqref{grav4c}\nonumber\\
&=(-\scalG/2\scalh^2)\{\gdiv{\vectr}{(\vecth/\scalr)}\}\nonumber\\
&=(-\scalG/2\scalh^2)\{(1/\scalr)[\gdiv{\vectr}{\vecth}]+\vecth[\gdiv{\vectr}{(1/\scalr)}]\}\nonumber\\
&=(-\scalG/2\scalh^2)\{\vecth[\dprod{\vectr}{\grad{(1/\scalr)}}]\}\nonumber\\
&=\scalG\vecth/2\scalr\scalh^2\nonumber\\
&=-\vectOme\beqref{grav4c}
\end{align}
\end{subequations}
which we shall found useful later. Substituting \eqnref{grav7} into \eqnref{grav6}
yields
\begin{equation}\label{grav9}
\curl{\vectb}
=-\gdiv{(\curl{\vectB})}{\vectu} + \udiv{(\curl{\vectB})}-2\curl{\biggl\{
  \frac{(\dprod{\vectB}{\vectr})(\cprod{\vectr}{\vectOme})}{\scalr^2}
   \biggr\}}.
\end{equation}
The last term of this equation is expanded as
\begin{align}\label{grav10}
\curl{\biggl\{\frac{(\dprod{\vectB}{\vectr})
  (\cprod{\vectr}{\vectOme})}{\scalr^2}\biggr\}}
&=\cprod{[\grad{(\dprod{\vectB}{\vectr}/\scalr^2)}]}{(\cprod{\vectr}{\vectOme})}
  +(\dprod{\vectB}{\vectr}/\scalr^2)[\curl{(\cprod{\vectr}{\vectOme})}]\nonumber\\
\begin{split}
&=-\cprod{(\cprod{\vectr}{\vectOme})}{[\grad{(\dprod{\vectB}{\vectr}/\scalr^2)}]}
  +(\dprod{\vectB}{\vectr}/\scalr^2)[\vectr(\dive{\vectOme})\\
&\qquad-\gdiv{\vectr}{\vectOme}
  +\gdiv{\vectOme}{\vectr}-\vectOme(\dive{\vectr})]\beqref{vid3}
\end{split}
\nonumber\\
&=-\cprod{(\cprod{\vectr}{\vectOme})}{[\grad{(\dprod{\vectB}{\vectr}/\scalr^2)}]}
  -\vectOme(\dprod{\vectB}{\vectr}/\scalr^2)\beqref{grav4b}\ \&\ \eqnref{grav8b}\nonumber\\
&=-\frac{\vectOme(\dprod{\vectB}{\vectr})}{\scalr^2}
-\cprod{(\cprod{\vectr}{\vectOme})}{\biggl\{
  \frac{1}{\scalr^2}\grad{(\dprod{\vectB}{\vectr})}+(\dprod{\vectB}{\vectr})
  \grad{\biggl(\frac{1}{\scalr^2}\biggr)}\biggr\}}\nonumber\\
&=-\frac{\vectOme(\dprod{\vectB}{\vectr})}{\scalr^2}+\frac{2(\dprod{\vectB}{\vectr})
  [\cprod{(\cprod{\vectr}{\vectOme})}{\vectr}]}{\scalr^4}-
  \cprod{\frac{(\cprod{\vectr}{\vectOme})}{\scalr^2}}
  {\grad{(\dprod{\vectB}{\vectr})}}\nonumber\\
&=\frac{\dprod{\vectB}{\vectr}}{\scalr^2}\biggl\{-\vectOme
  -\frac{2\cprod{\vectr}{(\cprod{\vectr}{\vectOme})}}{\scalr^2}\biggr\}
  -\cprod{\frac{(\cprod{\vectr}{\vectOme})}{\scalr^2}}
  {\grad{(\dprod{\vectB}{\vectr})}}\nonumber\\
&=\frac{\vectOme(\dprod{\vectB}{\vectr})}{\scalr^2}
  -\cprod{\frac{(\cprod{\vectr}{\vectOme})}{\scalr^2}}
  {\grad{(\dprod{\vectB}{\vectr})}}\beqref{grav4c}\nonumber\\
\begin{split}
&=\frac{\vectOme(\dprod{\vectB}{\vectr})}{\scalr^2}
  -\cprod{\frac{(\cprod{\vectr}{\vectOme})}{\scalr^2}}{}\biggl\{
  \cprod{\vectB}{(\curl{\vectr})}\\
  &\qquad+\gdiv{\vectB}{\vectr}+\gdiv{\vectr}{\vectB}
  +\cprod{\vectr}{(\curl{\vectB})}\biggr\}\beqref{vid2}
\end{split}
\nonumber\\
&=\frac{\vectOme(\dprod{\vectB}{\vectr})}{\scalr^2}
  +\frac{\cprod{\vectB}{(\cprod{\vectr}{\vectOme})}}{\scalr^2}
  -\cprod{\frac{(\cprod{\vectr}{\vectOme})}{\scalr^2}}{[\gdiv{\vectr}{\vectB}]}
  -\cprod{\frac{(\cprod{\vectr}{\vectOme})}{\scalr^2}}{[\cprod{\vectr}{(\curl{\vectB})}]}
\nonumber\\
&=\frac{\vectr(\dprod{\vectB}{\vectOme})}{\scalr^2}
  -\frac{\cprod{(\cprod{\vectr}{\vectOme})}{[\gdiv{\vectr}{\vectB}]}}{\scalr^2}
  -\frac{\cprod{(\cprod{\vectr}{\vectOme})}[\cprod{\vectr}{(\curl{\vectB})}]}{\scalr^2}.
\end{align}
Substituting \eqnref{grav10} into \eqnref{grav9} yields
\begin{equation}\label{grav11}
\begin{split}
\curl{\vectb}
&=-\gdiv{(\curl{\vectB})}{\vectu} + \udiv{(\curl{\vectB})}
-\frac{2\vectr(\dprod{\vectB}{\vectOme})}{\scalr^2}\\
  &\qquad+\frac{2\cprod{(\cprod{\vectr}{\vectOme})}{[\gdiv{\vectr}{\vectB}]}}{\scalr^2}
  +\frac{2\cprod{(\cprod{\vectr}{\vectOme})}[\cprod{\vectr}{(\curl{\vectB})}]}{\scalr^2}.
\end{split}
\end{equation}
Again, our goal is to obtain an expression from $\curl{\vectb}$ that contains only
$\curl{\vectB}$. We can achieve this by multiplying both sides scalarwise with
$\cprod{\vectr}{\vectOme}$ so that the last three terms vanish and we get
\begin{align}\label{grav12}
\dprod{(\cprod{\vectr}{\vectOme})}{(\curl{\vectb})}
&=\dprod{(\cprod{\vectr}{\vectOme})}{[\udiv{(\curl{\vectB})}]}
-\dprod{(\cprod{\vectr}{\vectOme})}{[\gdiv{(\curl{\vectB})}{\vectu}]}\nonumber\\
\begin{split}
&=\dprod{(\cprod{\vectr}{\vectOme})}{[\udiv{(\curl{\vectB})}]}\nonumber\\
&\qquad-\dprod{2(\cprod{\vectr}{\vectOme})}{
[\scalr^{-2}(\dprod{(\curl{\vectB})}{\vectr})(\cprod{\vectr}{\vectOme})
  -\cprod{(\curl{\vectB})}{\vectOme}]}\beqref{grav8c}
\end{split}
\nonumber\\
\begin{split}
&=\dprod{(\cprod{\vectr}{\vectOme})}{[\udiv{(\curl{\vectB})}]}
-2\scalr^{-2}[\dprod{\vectr}{(\curl{\vectB})}][(\cprod{\vectr}{\vectOme})^2]\nonumber\\
&\qquad+2\bigl\{\dprod{(\cprod{\vectr}{\vectOme})}{[\cprod{(\curl{\vectB})}{\vectOme}]}\bigr\}
\end{split}
\nonumber\\
\begin{split}
&=\dprod{(\cprod{\vectr}{\vectOme})}{[\udiv{(\curl{\vectB})}]}
-2\scalr^{-2}[\dprod{\vectr}{(\curl{\vectB})}][\scalr^2\Omerep^2-(\dprod{\vectr}{\vectOme})^2]\nonumber\\
&\qquad+2\bigl\{\Omerep^2[\dprod{\vectr}{(\curl{\vectB})}]-
  (\dprod{\vectr}{\vectOme})[\dprod{\vectOme}{(\curl{\vectB})}]\bigr\}
  \beqref{vid9}
\end{split}
\nonumber\\
\begin{split}
&=\dprod{(\cprod{\vectr}{\vectOme})}{[\udiv{(\curl{\vectB})}]}
+2\scalr^{-2}(\dprod{\vectr}{\vectOme})^2[\dprod{\vectr}{(\curl{\vectB})}]\nonumber\\
&\qquad-2(\dprod{\vectr}{\vectOme})[\dprod{\vectOme}{(\curl{\vectB})}]
\end{split}
\nonumber\\
&=\dprod{(\cprod{\vectr}{\vectOme})}{[\udiv{(\curl{\vectB})}]}\beqref{grav4c}\nonumber\\
\begin{split}
&=\murep\dprod{(\cprod{\vectr}{\vectOme})}{[\udiv{(\fideriv{\vectD})}]}\\
  &\qquad+\murep\dprod{(\cprod{\vectr}{\vectOme})}{[\udiv{(\udiv{\vectD})}]}
\beqref{curlb}\ \&\ \eqnref{EM:D3}.
\end{split}
\end{align}

\subsection{The Wave Equation}
From \eqnref{WVEQ} we have
\begin{align}\label{grav13}
c^2[\lapl{\vectD}-\grad{(\dive{\vectD})}]
&=\sideriv{\vectD}{2}+\fideriv{\vectj}+(1/\murep)(\curl{\vectb})\nonumber\\
&=\sideriv{\vectD}{2}+\udiv{(\fideriv{\vectD})}+(1/\murep)(\curl{\vectb})\beqref{grav5}.
\end{align}
Multiplying both sides of this equation scalarwise by $\cprod{\vectr}{\vectOme}$
yields
\begin{align}\label{grav14a}
\begin{split}
c^2\dprod{(\cprod{\vectr}{\vectOme})}{[\lapl{\vectD}-\grad{(\dive{\vectD})}]}
&=\dprod{(\cprod{\vectr}{\vectOme})}{[\sideriv{\vectD}{2}]}+
  \dprod{(\cprod{\vectr}{\vectOme})}{[\udiv{(\fideriv{\vectD})}]}\\
  &\qquad+(1/\murep)[\dprod{(\cprod{\vectr}{\vectOme})}{(\curl{\vectb})}]
\end{split}
\end{align}
which, in view of \eqnref{grav12}, leads to the wave equation
\begin{align}\label{grav14b}
\begin{split}
c^2\dprod{(\cprod{\vectr}{\vectOme})}{[\lapl{\vectD}-\grad{(\dive{\vectD})}]}
&=\dprod{(\cprod{\vectr}{\vectOme})}{[\sideriv{\vectD}{2}]}+
  2\dprod{(\cprod{\vectr}{\vectOme})}{[\udiv{(\fideriv{\vectD})}]}\\
  &\qquad+\dprod{(\cprod{\vectr}{\vectOme})}{[\udiv{(\udiv{\vectD})}]}.
\end{split}
\end{align}
The last term of this equation can be evaluated for
the plane wave of \eqnref{PLNWV} as
\begin{align}\label{grav15}
\udiv{[\udiv{\vectD}]}
&=\udiv{[(\dprod{\vectu}{i\vectkap})\vectD]}\nonumber\\
&=i(\dprod{\vectu}{\vectkap})[\udiv{\vectD}]+i\vectD[\udiv{(\dprod{\vectu}{\vectkap})}]\nonumber\\
&=i(\dprod{\vectu}{\vectkap})[(\dprod{\vectu}{i\vectkap})\vectD]
  +i\vectD[\dprod{\vectu}{\grad{(\dprod{\vectu}{\vectkap})}}]\nonumber\\
\begin{split}
&=-(\dprod{\vectu}{\vectkap})^2\vectD
 +i\vectD[\dprod{\vectu}{}\{\cprod{\vectu}{(\curl{\vectkap})}\\
   &\qquad+\udiv{\vectkap}+\gdiv{\vectkap}{\vectu}+\cprod{\vectkap}{(\curl{\vectu})\}}]
   \beqref{vid2}
\end{split}
\nonumber\\
&=-(\dprod{\vectu}{\vectkap})^2\vectD+i\vectD[\dprod{\vectu}{\{2\cprod{\vectkap}{\vectOme}
  +\gdiv{\vectkap}{\vectu}\}}]\beqref{grav4c}\nonumber\\
\begin{split}
&=-(\dprod{\vectu}{\vectkap})^2\vectD+2i\vectD[\dprod{\vectu}{(\cprod{\vectkap}{\vectOme})}]\\
  &\qquad+2i\vectD\biggl[\dprod{\vectu}{\biggl\{\frac{(\dprod{\vectkap}{\vectr})(\cprod{\vectr}
  {\vectOme})}{\scalr^2}-\cprod{\vectkap}{\vectOme}\biggr\}}\biggr]\beqref{grav8a}
\end{split}
\nonumber\\
&=-(\dprod{\vectu}{\vectkap})^2\vectD+2i\scalr^{-2}(\dprod{\vectkap}{\vectr})
  [\dprod{\vectu}{(\cprod{\vectr}{\vectOme})}]\vectD\nonumber\\
&=-(\dprod{\vectu}{\vectkap})^2\vectD+2i\scalr^{-2}(\dprod{\vectkap}{\vectr})
  [\dprod{\vectOme}{(\cprod{\vectu}{\vectr})}]\vectD\nonumber\\
&=-(\dprod{\vectu}{\vectkap})^2\vectD+2i\scalr^{-2}(\dprod{\vectkap}{\vectr})
  (\dprod{\vectOme}{\vecth})\vectD\beqref{grav2b}\nonumber\\
&=-(\dprod{\vectu}{\vectkap})^2\vectD-i\scalG\scalr^{-3}(\dprod{\vectkap}{\vectr})\vectD
  \beqref{grav4c}\nonumber\\
&=-(\dprod{\vectu}{\vectkap})^2\vectD+i(\dprod{\vectkap}{\vectg})\vectD\beqref{grav1}.
\end{align}
Substituting \eqnref{nabids} and \eqnref{grav15} into \eqnref{grav14b} leads to the dispersion
equation
\begin{equation}\label{grav16}
[\dprod{\vectD}{(\cprod{\vectr}{\vectOme})}][\psirep^2-c^2\kaprep^2
  -i(\dprod{\vectkap}{\vectg})]
  +c^2(\dprod{\vectkap}{\vectD})[\dprod{\vectkap}{(\cprod{\vectr}{\vectOme})}]=0.
\end{equation}
If $\vectD, \vectr\text{ and }\vectOme$ are coplanar, the dispersion equation becomes
$[\dprod{\vectkap}{\vectD}][\dprod{\vectkap}{(\cprod{\vectr}{\vectOme})}]=0$ which
describes a degenerate or nonpropagating field.
To obtain a nondegenerate solution, we shall require that
$\dprod{\vectD}{(\cprod{\vectr}{\vectOme})}\ne0$.
In this case the dispersion equation can be rewritten as
\begin{subequations}\label{grav17}
\begin{equation}\label{grav17a}
\psirep^2+c^2(\tmpa^2-\kaprep^2)-i\alprep=0
\end{equation}
where
\begin{equation}\label{grav17b}
\alprep=\dprod{\vectkap}{\vectg},\quad
\tmpa^2=(\dprod{\vectkap}{\vectD})\biggl\{\frac{\dprod{\vectkap}{(\cprod{\vectr}{\vectOme})}}
  {\dprod{\vectD}{(\cprod{\vectr}{\vectOme})}}\biggr\}
=(\dprod{\vectkap}{\vectD})\biggl\{\frac{\dprod{\vectkap}{(\cprod{\vectr}{\vecth})}}
  {\dprod{\vectD}{(\cprod{\vectr}{\vecth})}}\biggr\}.
\end{equation}
\end{subequations}

\subsection{Eccentric Waves ($\tmpa=\kaprep$)}
For these waves, we obtain from \eqnref{abbrv2} and \eqnref{grav17a} the system of equations
\begin{subequations}\label{grav18}
\begin{equation}\label{grav18a}
\rpsi^2-\ipsi^2=0
\end{equation}
\begin{equation}\label{grav18b}
2\rpsi\ipsi=\alprep
\end{equation}
\end{subequations}
with solutions
\begin{subequations}
\begin{equation}
\rpsi = \pm\kaprep\etarep,\quad\ipsi=\pm\kaprep\etarep
\qquad(\dprod{\vectkap}{\vectg}\ge0)
\end{equation}
\begin{equation}
\rpsi = \pm\kaprep\etarep,\quad\ipsi=\mp\kaprep\etarep
\qquad(\dprod{\vectkap}{\vectg}\le0)
\end{equation}
where, as usual, all the upper signs or all the lower signs are to be taken
together in each set of equations, and $\etarep\ge0$ satisfies
\begin{equation}
\etarep^2 = \biggl|\frac{\alprep}{2\kaprep^2}\biggr|.
\end{equation}
\end{subequations}
The corresponding frequency components is given by \eqnref{abbrv3} as
\begin{subequations}\label{grav19}
\begin{equation}\label{grav19a}
\rome=\pm\kaprep\etarep - \dprod{\vectu}{\vectkap},\quad\iome=\pm\kaprep\etarep
\qquad(\dprod{\vectkap}{\vectg}\ge0)
\end{equation}
\begin{equation}\label{grav19b}
\rome=\pm\kaprep\etarep - \dprod{\vectu}{\vectkap},\quad\iome=\mp\kaprep\etarep
\qquad(\dprod{\vectkap}{\vectg}\le0).
\end{equation}
\end{subequations}
Substituting these into \eqnref{wvprps} gives
\begin{subequations}\label{grav20}
\begin{equation}\label{grav20a}
\left.
\begin{aligned}
\phvel &= (\pm\etarep - u\cos\thtrep)\unitkap\\
\grvel &= \pm\vectg/(4\kaprep\etarep) - \vectu \\
\sklife &= \pm(\kaprep\etarep)^{-1}
\end{aligned}
\qquad\right\}(\dprod{\vectkap}{\vectg}>0)
\end{equation}
\begin{equation}\label{grav20b}
\left.
\begin{aligned}
\phvel &= (\pm\etarep - u\cos\thtrep)\unitkap\\
\grvel &= \pm\vectg/(4\kaprep\etarep) - \vectu \\
\sklife &= \mp(\kaprep\etarep)^{-1}
\end{aligned}
\qquad\right\}(\dprod{\vectkap}{\vectg}<0)
\end{equation}
\begin{equation}\label{grav20c}
\left.
\begin{aligned}
\phvel &= -u\cos\thtrep\unitkap\\
\grvel &= -\vectu\\
\sklife &= \infty.
\end{aligned}
\qquad\right\}(\dprod{\vectkap}{\vectg}=0)
\end{equation}
\end{subequations}

\subsection{Noneccentric Waves ($\tmpa\ne\kaprep$)}
From \eqnref{abbrv2} and \eqnref{grav17a}, we obtain for these waves the system of equations
\begin{subequations}\label{grav21}
\begin{equation}\label{grav21a}
\rpsi^2-\ipsi^2=c^2(\kaprep^2-\tmpa^2)
\end{equation}
\begin{equation}\label{grav21b}
2\rpsi\ipsi=\alprep
\end{equation}
\end{subequations}
with solutions
\begin{subequations}\label{grav22}
\begin{equation}\label{grav22a}
\rpsi=\pm c\dragf\kaprep,\quad\ipsi=\pm c\kaprep\sqrt{\dragf^2-\gamrep^2}
\qquad(\dprod{\vectkap}{\vectg}\ge0)
\end{equation}
\begin{equation}\label{grav22b}
\rpsi=\pm c\dragf\kaprep,\quad\ipsi=\mp c\kaprep\sqrt{\dragf^2-\gamrep^2}
\qquad(\dprod{\vectkap}{\vectg}\le0)
\end{equation}
\end{subequations}
where $\dragf>0$ satisfies
\begin{equation}\label{grav23}
\dragf=\gamrep\left\{\frac{1+\sqrt{1+\vthtrep^2}}{2}\right\}^{1/2},\quad
\gamrep^2=\left|1-\frac{\tmpa^2}{\kaprep^2}\right|,\quad
\vthtrep=\frac{\alprep}{c^2\kaprep^2\gamrep^2}.
\end{equation}
The corresponding frequency components are given by \eqnref{abbrv3} as
\begin{subequations}\label{grav24}
\begin{equation}\label{grav24a}
\rome=\pm c\dragf\kaprep-\dprod{\vectu}{\vectkap},
\quad\iome=\pm c\kaprep\sqrt{\dragf^2-\gamrep^2}
\qquad(\dprod{\vectkap}{\vectg}\ge0)
\end{equation}
\begin{equation}\label{grav24b}
\rome=\pm c\dragf\kaprep-\dprod{\vectu}{\vectkap},
\quad\iome=\mp c\kaprep\sqrt{\dragf^2-\gamrep^2}
\qquad(\dprod{\vectkap}{\vectg}\le0).
\end{equation}
\end{subequations}
Substituting these into \eqnref{wvprps} gives
\begin{subequations}\label{grav25}
\begin{equation}\label{grav25a}
\left.
\begin{aligned}
\phvel &= (\pm c\dragf - u\cos\thtrep)\unitkap\\
\grvel &= \pm(c\dragf\unitkap + \vectw) - \vectu\\
\sklife &= \pm\left(c\kaprep\sqrt{\dragf^2-\gamrep^2}\right)^{-1}
\end{aligned}
\qquad\right\}(\dprod{\vectkap}{\vectg}\ge0)
\end{equation}
\begin{equation}\label{grav25b}
\left.
\begin{aligned}
\phvel &= (\pm c\dragf - u\cos\thtrep)\unitkap\\
\grvel &= \pm(c\dragf\unitkap + \vectw) - \vectu\\
\sklife &= \mp\left(c\kaprep\sqrt{\dragf^2-\gamrep^2}\right)^{-1}
\end{aligned}
\qquad\right\}(\dprod{\vectkap}{\vectg}\le0)
\end{equation}
where
\begin{equation}
\vectw=\frac{\chirep}{4c\kaprep\dragf}
\biggl\{\vectg - \vpirep_1\vectkap\biggr\}
+\vpirep_2(\cprod{\vectr}{\vecth})+\vpirep_3\vectD 
\end{equation}
and
\begin{align}
\begin{split}
&\vpirep_1=\frac{2}{\kaprep^2}\left\{
  \alprep-\frac{2c\kaprep\dragf\betrep\tmpa^2\Gamrep}{\chirep}\right\},\quad
\vpirep_2=\frac{\Gamrep(\dprod{\vectkap}{\vectD})}{2},\quad
\vpirep_3=\frac{\Gamrep(\dprod{\vectkap}{(\cprod{\vectr}{\vecth})})}{2}\\
&\betrep = \dprod{\vectD}{(\cprod{\vectr}{\vecth})},\quad
\chirep = \vthtrep(1+\vthtrep^2)^{-1/2},\quad
\Gamrep=\frac{\alprep\chirep-2c^2\kaprep^2\dragf^2}
  {2\dragf\gamrep^2c\kaprep^3\betrep}.
\end{split}
\end{align}
\end{subequations}

\section{Applications}\label{APPS}
We have shown that for an observer in translational, rotational or gravitational
motion, a linearly polarized plane wave has two modes of propagation as
summarized in \tabref{TAB1}. In the regular mode, the phase and group velocities
of the wave depend not only on wavelength and the observer's acceleration but also
(via $c$ as defined by \eqnref{WVEQ}) on the permittivity and the permeability of a
medium. In the irregular or heretic mode, the phase
and group velocities of the wave depend only on wavelength and the observer's motion
but not on the permittivity or permeability of the medium. A direct consequence of this
circumstance is that phenomena (such as the Sagnac effect) which result in small 
measurable quantities due
to the largeness of $c$ can be made much more accurate by using heretic waves instead of
regular waves. 
\begin{table}[htb]
\begin{center}
\begin{tabular}{|c|c|c|}
\hline
                  & \multicolumn{2}{c|}{mode}\\
\cline{2-3}
motion            & regular               & heretic\\
\hline\hline
                  & transverse            & nontransverse\\
translation       & \eqnref{trans12}       & \eqnref{trans16}\\
\hline
                  & noncoplanar           & coplanar\\
rotation          & \eqnref{rota28}        & \eqnref{rota22}\\
\hline
                  & noneccentric          & eccentric\\
gravitation       & \eqnref{grav25}        & \eqnref{grav20}\\
\hline
\end{tabular}
\end{center}
\caption{Propagation modes of a linearly polarized plane wave for
an accelerated observer. The equation describing each mode in this paper
is shown in parentheses.}
\label{TAB1}
\end{table}
Another consequence is that heretic waves radiated with a given speed
are observable only by an observer moving with the same speed, hence they are well suited
for selective broadcasting. They can be used, for example, to control speed on a highway
if a transmitter is installed on the highway and a receiver is installed in a car
because when (and only when) the car reaches the speed limit, the receiver will 
pick up the heretic wave from the transmitter and take
appropriate action. Inasmuch as for an observer moving with a given velocity there exists
not one heretic wave with a definite frequency and wavelength but a complete spectrum of
heretic waves with different frequencies and wavelengths, all propagating with 
the same speed as the observer, other uses of heretic waves based on their 
frequency and wavelength are conceivable.

It is easy to see that a plane wave that is linearly polarized in its direction
of propagation will show hereticity relative to an observer 
in translational,
rotational or gravitational motion. For if \vectD\ and $\vectkap$ are collinear,
we have $\dprod{\vectkap}{\vectD}\ne0$ in \eqnref{trans6} for a translating observer,
$\betrep=0$ in \eqnref{rota17} for a
rotating observer, and $\tmpa=\kaprep$ in \eqnref{grav17} for a gravitating
observer. In the same way, it is easy to see that if the wave is linearly polarized
at right angles to its direction of propagation, it will behave in a regular way.
This implies that if the angle between the polarization and the propagation direction
is varied from $\pi/2$ to $0$, the wave speed will vary from a value controlled
by the material constants of the medium to a value controlled by the motion of
the observer. Hence if the observer is at rest, the wave will effectively be stopped 
from propagating. Propagation can be of course be resumed
by varying the polarization angle in the opposite direction; that is, from $0$ to 
$\pi/2$. 

We have shown that waves polarized at right angles to the wave vector can be 
heretic relative a rotating observer if \eqnref{rota30} is satisfied. 
This result provides an interesting
explanation of the apparently spurious signals observed in the experiment of
Brillet and Hall~\cite{Brillet79}. To see this, suppose that \vectD\ and $\vectkap$ are on
the horizontal plane so that $\cprod{\vectkap}{\vectD}$ is vertical.
Equation \eqnref{rota30} shows that when the table used in the experiment is 
very nearly horizontal so that $\thtrep_3$ is close to 0, both $\thtrep_1$ and 
$\thtrep_2$ approach $\pi/2$.
Hence $\pfreq$ approaches zero and no spurious signal should be observed. 
As the rotation axis of the table deviates from the vertical in such a way that
$\thtrep_1\ne\pi/2$ and $\thtrep_2\ne\pi/2$, $\thtrep_3$ approaches 
$\pi/2$. Hence $\sec\thtrep_3$ and $\pfreq$ become increasingly large. This
explains why the experimenters were able to reduce the frequency of the spurious signal
from 1KHz to 17Hz by increasingly aligning the rotation axis of the table 
vertically. Equation \eqnref{rota30} also shows that for a given alignment
of the rotation axis, the signal should have an amplitude $\approx 2\Omerep$. This
means that the signal should peak twice for each rotation of the table, as was
observed in the experiment. It appears therefore that Brillet and Hall may have unknowingly 
detected heretic waves. But it is not clear how the larger observed signal of $\approx$ 200Hz 
can be explained by heretic waves. For this and other reasons, the evidence is not as conclusive 
as one would like, and the spurious signals may very well be due to other 
causes~\cite{Aspden81, Klauber00a}.

Except for the situation described by \eqnref{rota30}, heretic modes of propagation 
arise mainly because we have assumed a nonzero charge 
density in the region of interest ($\sigrep\ne0$). As a result, they may be less
convenient than regular waves for some applications. One area where regular waves
have found common use in industry is in the detection of acceleration. Our analysis
of regular waves can be used to measure acceleration in three ways. First, one can
use the equations for the phase velocity to calculate the number $N$ of
fringes obtainable from the interference of regular waves. Since this number
depends on $\dragf$ which in turn depends on the component of the acceleration in
the direction of propagation via $\vthtrep$, this component of the acceleration can be 
calculated 
from a measurement of $N$. Second, one can use the equations for the group velocity
of a regular wave to calculate the aberration $\vphirep$ of the wave. The aberration 
will be found to depend on $\dragf$ so that the component of the acceleration in
the propagation direction can be calculated from a measurement of $\vphirep$.
Finally, one can measure the skin time $\sklife$ directly and calculate the relevant
acceleration component from the dependence of $\sklife$ on the component.
To illustrate the $\sklife$ method for translation, we observe that
to a second order approximation in $\vthtrep$, \eqnref{trans10x} gives
\begin{equation*}
\dragf^2 - 1 = \vthtrep^2/4
\end{equation*}
while from \eqnref{trans12a}, we have
\begin{equation*}
\dragf^2 - 1 = (c\kaprep\sklife)^{-2}
\end{equation*}
and from \eqnref{trans10x}, to all orders of accuracy,
\begin{equation*}
\vthtrep = \scalap/(c^2\kaprep)
\end{equation*}
where $\scalap$ is the component of the acceleration \vecta\ parallel to $\vectkap$.
It follows from these equations that
\begin{equation*}
\scalap=\frac{2c}{\sklife}
\end{equation*}
which allows $\scalap$ to be calculated from a measurement of $\sklife$. As is usually
done in practice with other methods, the components of \vecta\ along three orthogonal 
directions can be measured by using a system of three waves propagating in these directions. 

For an observer in uniform translational motion, it appears that one should be able to
use the phase velocity in \eqnref{trans12c} to calculate the number $N$ of fringes involved
in the interference of two regular waves, and that, by measuring $N$ and using its dependence
on the observer's speed, the speed can be estimated. This is indeed correct. Unfortunately,
as is well known,
attempts to use this technique with the Michelson-Morley interferometer have failed
to date. This circumstance is interpreted in special relativity as indicating the
incorrectness of the phase velocity in \eqnref{trans12c}. We are compelled to interprete 
the situation differently, however, because the phase velocity in question follows 
unavoidably from our field equations, and our field equations have been shown to be 
deductive consequences of conventions~\cite{Adewole00a, Adewole01a}, which means that
they cannot be falsified by any experiment consistent with those 
conventions~(\cite{Eddington26}, pg~214). This problem will be treated in detail
elsewhere. We conclude with the remark that the results obtained in this paper 
have been shown to be valid for linearly polarized plane waves and may not be applied 
to nonplane, unpolarized or nonlinearly polarized waves without further analysis.
They may also not be applied without further analysis in a medium that is moving, 
anisotropic or inhomogeneous, or in a region of space where the divergence theorem 
does not hold or charge conservation is violated. For a comparative study of our
treatment with others,
we refer the reader to the works of Phipps~\cite{Phipps00}, Klauber~\cite{Klauber00b}, 
Goy and Selleri~\cite{Goy97}, Rizzi and Tartaglia~\cite{Rizzi98}, L\"ammerzahl and 
Haugan~\cite{Lammerzahl01}, Kopeikin and Sch\"afer~\cite{Kopeikin99},
 and the references contained in these works.

\appendix
\section{Appendix: Vector Identities}
The following vector identities are used in this paper
(see, for example,~\cite{Tromba88}). For any analytic vectors 
\vecte, \vectf, \vectg, \vecth\ and scalar $\phirep$,
\begin{equation}\label{vid4}
\curl{(\grad{\phirep})} = \vectzero
\end{equation}
\begin{equation}\label{vid5}
\dive{(\curl{\vectf})} = 0
\end{equation}
\begin{equation}\label{vid1}
\curl{(\curl{\vectf})} = \grad{(\dive{\vectf})} - \lapl{\vectf}
\end{equation}
\begin{equation}\label{vid6}
\curl{(\phirep\vectf)} = \phirep(\curl{\vectf}) - 
\cprod{\vectf}{(\grad{\phirep})}
\end{equation}
\begin{equation}\label{vid8}
\gdiv{\vectf}{(\phirep\vectg)}=
\vectg(\dprod{\vectf}{\grad{\phirep}}) +
\phirep(\gdiv{\vectf}{\vectg})
\end{equation}
\begin{equation}\label{vid7}
\gdiv{\vectf}{(\cprod{\vectg}{\vecth})}=
\cprod{\vectg}{(\gdiv{\vectf}{\vecth})} -
\cprod{\vecth}{(\gdiv{\vectf}{\vectg})}
\end{equation}
\begin{equation}\label{vid3}
\curl{(\cprod{\vectf}{\vectg})} = \vectf(\dive{\vectg}) -
\gdiv{\vectf}{\vectg} + \gdiv{\vectg}{\vectf} -
\vectg(\dive{\vectf})
\end{equation}
\begin{equation}\label{vid2}
\grad{(\dprod{\vectf}{\vectg})} = \cprod{\vectf}{(\curl{\vectg})}
+ \gdiv{\vectf}{\vectg} + \gdiv{\vectg}{\vectf} +
\cprod{\vectg}{(\curl{\vectf})}
\end{equation}
\begin{equation}\label{vid9}
\dprod{(\cprod{\vecte}{\vectf})}{(\cprod{\vectg}{\vecth})}
=(\dprod{\vecte}{\vectg})(\dprod{\vectf}{\vecth})
-(\dprod{\vecte}{\vecth})(\dprod{\vectf}{\vectg}).
\end{equation}

\bibliographystyle{unsrt}
\bibliography{xbib}

\section*{Errata 22apr2001}
\begin{itemize}
\item sign correction in \eqnref{abbrv4}, no further consequences
\item sign correction in \eqnref{grav2}, minor consequences for some equations
\item constant factor correction in \eqnref{grav4b}, no further consequences
\item added \eqnref{vid9} and noted its use in \eqnref{grav12}
\end{itemize}

\section*{Errata 07may2001}
\begin{itemize}
\item added footnotes to clarify \eqnref{EM:B1} and \eqnref{rota29}
\item minor corrections in spelling, grammar and symbols
\item changed position on the 200Hz spurious signal in the experiment of Brillet and Hall
\end{itemize}
\end{document}